\documentclass[preprint,amsmath,amssymb]{revtex4}

\usepackage{latexsym} 
\usepackage{amssymb}  
\usepackage{amsbsy}   
\usepackage{graphicx}       
\usepackage[dvips]{color}

\newcommand{\bm}[1]{\mbox{\boldmath${#1}$}}

\newcommand{\al}{\alpha}

\newcommand{\Ga}{\Gamma}
\newcommand{\de}{\delta}
\newcommand{\De}{\Delta}

\newcommand{\eps}{\epsilon}

\newcommand{\ka}{\kappa}
\newcommand{\la}{\lambda}

\newcommand{\Si}{\Sigma}
\renewcommand{\th}{\theta}   

\newcommand{\p}{\partial}
 
\newcommand{\txt}{\textstyle}

\newcommand\eqn[1]{(\ref{#1})}      
\newcommand\Eqn[1]{Eq.~(\ref{#1})}  

\newcommand{\beq}{\begin{equation}}
\newcommand{\eeq}{\end{equation}}
\newcommand{\ba}{\begin{array}}
\newcommand{\bea}{\begin{eqnarray}}
\newcommand{\ea}{\end{array}}
\newcommand{\eea}{\end{eqnarray}}
\newcommand{\bi}{\begin{itemize}}  
\newcommand{\ei}{\end{itemize}}
\newcommand{\ben}{\begin{enumerate}} 
\newcommand{\een}{\end{enumerate}}
\newcommand{\bc}{\begin{center}}
\newcommand{\ec}{\end{center}}
\newcommand{\bl}{\begin{flushleft}}
\newcommand{\el}{\end{flushleft}}
\newcommand{\br}{\begin{flushright}}
\newcommand{\er}{\end{flushright}}

\newcommand\comment[1]{ \hbox{[{\it Comment suppressed here.}\/]} }
\newcommand\hide[1]{}
\newcommand\deriv[2]{\frac{d{#1}}{d{#2}}}
\newcommand\pderiv[2]{\frac{\p{#1}}{\p{#2}}}

\newcommand{\Tr}{\hbox{Tr}}

\newcommand{\skipover}[1]{}
\newcommand{\nn}{\nonumber \\}

\newcommand{\half} {{\txt \frac{1}{2}}}

\numberwithin{equation}{section}
\renewcommand{\theequation}{\arabic{section}.\arabic{equation}}

\newcommand{\MeV}{{\rm MeV}}

\newcommand{\f}{f_p^0}

\begin{document}
\title{Thermal conductivity of color-flavor locked quark matter}

\author{Matt Braby, Jingyi Chao and Thomas Sch\"afer}
\address{Physics Department \\
North Carolina State University \\
Raleigh, NC 27695, USA}

\date{\today}

\begin{titlepage}
\renewcommand{\thepage}{}          

\begin{abstract}
We compute the thermal conductivity of color-flavor locked (CFL) 
quark matter. At temperatures below the scale set 
by the gap in the quark spectrum, transport properties are determined
by collective modes. In this work we focus on the contribution
from the lightest modes, the superfluid phonon and the massive 
neutral kaon. The calculation is done in the framework of kinetic 
theory, using variational solutions of the linearized Boltzmann 
equation.  We find that the thermal conductivity due to phonons is 
$\ka^{(P)} \sim 1.04 \times 10^{26}\ \mu_{500}^8\, \De_{50}^{-6}\ 
\rm{erg\ cm^{-1}\ s^{-1}\ K^{-1}}$ and the contribution of kaons is 
$\ka^{(K)} \sim 2.81 \times 10^{21}\ f_{\pi,100}^4 \, T_{\rm MeV}^{1/2}
\, m_{10}^{-5/2}\ \rm{erg\ cm^{-1}\ s^{-1}\ K^{-1}}$.  These values 
are smaller than previous estimates, but still much larger than (in 
the case of phonons) or similar to (for kaons) the corresponding values 
in nuclear matter. From the phonon thermal conductivity we estimate 
that a CFL quark matter core of a compact star becomes isothermal 
on a timescale of a few seconds.

\end{abstract}
\maketitle
\end{titlepage}

\section{Introduction}
In this paper, we explore the thermal conductivity of quark matter
in the regime of large baryon density and low temperature.  It is 
expected that cold dense quark matter is a color superconductor, 
and that at asymptotically high density the ground state of three-flavor
quark matter is the color-flavor locked (CFL) phase \cite{Alford:2007xm}.  
The work described in this paper is part of an ongoing research effort
with the goal of determining the transport properties of not only the 
CFL phase, but also of other, less dense, phases of quark matter. 
Previous work has focused on the shear viscosity \cite{Manuel:2004iv}, 
bulk viscosity \cite{Alford:2007rw,Manuel:2007pz}, and neutrino emissivity
of the CFL phase \cite{Jaikumar:2002vg,Reddy:2002xc}. A general discussion 
of the kinetics of a CFL superfluid is given in \cite{Mannarelli:2008jq}. 
There are also calculations of the transport properties of unpaired quark 
matter \cite{Iwamoto:1980eb,Heiselberg:1993cr,Schafer:2004jp}, 
and some results for the transport properties of a kaon condensed
CFL phase \cite{Alford:2008pb}. The long term goal is to connect 
these calculations of transport properties to possible observational 
signatures of high density quark matter phases in the core of compact 
stars.

The thermal conductivity of dense matter plays an important part in the 
cooling of compact stars \cite{Pethick:1991mk,Lattimer:1994,Yakovlev:2004iq}.
The cooling history of a star depends on the rate of energy loss
by neutrino emission from the bulk and, at very late times, photon
emission from the surface, on the specific heat, and on the thermal 
conductivity. Thermal conductivity determines how quickly different
layers of the star become isothermal. It is generally assumed that 
the core of the star becomes isothermal very quickly, but that the 
outer crust is a poor conductor of heat and may take several hundreds 
of years to reach the temperature of the inner core 
\cite{Shternin:2006uq,Horowitz:2008jt,Aguilera:2008ed}. In order to 
verify that this is true for CFL quark matter core we have to compute 
the thermal conductivity of CFL matter. There is an earlier calculation 
of the thermal conductivity of the CFL phase \cite{Shovkovy:2002kv}. 
This calculation was based on a simple mean free path estimate. 
In the present paper we will perform a more definitive calculation
based on the linearized Boltzmann equation. We will show that 
while the mean free path estimate is not reliable, the main 
conclusion of \cite{Shovkovy:2002kv} is valid -- a CFL quark
matter core becomes isothermal on a very short time scale. 

 The thermal conductivity, like other transport properties,
depends on the properties of quasi-particle excitations. In
the CFL phase up, down, and strange quarks are all gapped, with 
the gaps typically much larger than the appropriate temperature
of the compact stars.  Hence, the quarks are unlikely to have
any significant contribution to transport properties.  Below
the gap the excitation spectrum consists of collective modes
associated with the spontaneously broken symmetries of the 
CFL phase, see \cite{Casalbuoni:1999wu,Son:1999cm,Bedaque:2001je,Kryjevski:2004jw}.
There is a phonon mode related to superfluidity, which we can 
view as the Goldstone boson associated with the breaking of the 
$U(1)$ of baryon number, and there is a meson octet coming from 
the breaking of the $SU(3)_c \times SU(3)_L\times SU(3)_R$ color 
and chiral flavor symmetry to the diagonal subgroup $SU(3)_{C+F}$. 
This diagonal subgroup is the residual symmetry 
of the color-flavor locked diquark condensate. The quantum 
numbers of the meson octet coincide with those arising from 
the breaking of chiral symmetry in the QCD vacuum, and we 
will refer to these modes as pions, kaons, etc.  However, the mass 
hierarchy of the meson octet in CFL is different, \cite{Son:1999cm}, 
such that the kaons are the lightest excitations. In this paper, 
we will calculate the thermal conductivity due to the lightest
excitations in the CFL phase, phonons and kaons.  

This paper is organized as follows. In Section \ref{sec_transport}
and \ref{varsol} we will outline the calculation of the thermal 
conductivity in kinetic theory, in Section \ref{sec_coll} we will
discuss the interaction among the light excitations and derive 
the associated collision terms, and in Section \ref{sec_res} we
will present numerical results. We end with some conclusions in 
Section \ref{sec_concl}.

\section{Thermal Conductivity and Transport Theory}
\label{sec_transport}

The thermal conductivity is defined in hydrodynamics as 
\beq
{\bm q} = -\ka {\bm \nabla} T
\label{kappa_def}
\eeq
where ${\bf q}$ is the heat flow and $\ka$ is the thermal conductivity.
In kinetic theory, the heat flux can be written in terms of the 
quasi-particle distribution function as
\beq
{\bf q} = \int \frac{d^3 p}{(2\pi)^3} {\bf v_p} E_p \de f_p
\label{hf}
\eeq
%
%
where $v_p = \p E_p/\p p$ is the particle velocity and $\de f_p
=f_p-f_p^0$ is the deviation of the distribution function from 
the form in local thermal equilibrium, given by
\beq
f^0_p = \frac{1}{e^{(p_\mu u^\mu - \mu)/T} - 1}\, . 
\label{dist}
\eeq
Here, $u_\mu$ is the local fluid velocity and $\mu$ is the 
chemical potential (which is zero in the case of phonons and equal to the
hypercharge chemical potential in the case of the kaons). Since we are 
only interested in contributions to the stress-energy tensor arising from 
thermal gradients we can write $\de f_p$ as
\beq
\de f_p = -\frac{f_p^0 (1+f_p^0)}{T^3} g(p) 
  {\bm p} \cdot {\bm\nabla} T\, , 
\label{df}
\eeq
where $g(p)$ is a dimensionless function of the magnitude of the
momentum. Inserting this into \Eqn{hf}, we can then read off the heat flux as
\beq
{\bm q} = -\frac{{\bm \nabla}T}{3 T^3} \int \frac{d^3 p}{(2\pi)^3}
\, \f (1+\f) v_p\, E_p\, p\, g(p)
\label{heat}
\eeq
and the thermal conductivity as
\beq
\ka = \frac{1}{3 T^3} \int \frac{d^3 p}{(2\pi)^3}\, \f (1+\f) g(p)\, 
v_p\,E_p\, p
\label{kappa}
\eeq
Kinetic theory also gives some constrains on the form that $\de f_p$ 
can take. Conservation of particle number, energy and momentum 
imply \cite{Landau:kin}
\beq
\int d\Ga \de f_p = \int d\Ga E_p\, \de f_p 
     = \int d\Ga {\bm p}\, \de f_p = 0 \, , 
\label{del_f_constr}
\eeq
where $d\Ga = \frac{d^3 p}{(2\pi)^3}$.  Note that the
number of phonons is not explicitly conserved.  However,
we will only consider $2\leftrightarrow 2$ processes and
the particle number constraint is expected to hold.
We can show that for $\de f_p$ of the form given in \Eqn{df} 
the first and second constraints (energy and momentum conservation) 
are automatically satisfied, 
but the third one (particle number conservation) is non-trivial. We get 
\beq
0 = \int d\Ga\, {\bm p}\, \de f_p 
  = \frac{{\bm \nabla} T}{3\,T^3}\int d\Ga\,\f (1+\f) g(p)\,p^2 \, . 
\label{const3}
\eeq
Note that if $v_p$ is independent of $p$, then the 
contribution to the thermal conductivity vanishes due to the constraint. 
This implies that phonons with an exactly linear dispersion relation
do not contribute to the thermal conductivity. This result 
was first derived in connection with the transport properties 
of superfluid helium \cite{Khalatnikov:1965}. The thermal 
conductivity of superfluid helium is dominated by rotons
and phonon-roton scattering. Note that the phonon and roton
are part of the same excitation curve.  The name roton refers to the
part of the excitation curve in which the disperion relation is very
non-linear and $E(p)$ develops a second minimum.  In the following we will 
compute the thermal conductivity of the CFL phase due to
non-linearities in the phonon dispersion relation, and due 
to massive kaons. 

 To solve for the thermal conductivity, we need to determine 
the form of $g(p)$. The quasi-particle distribution functions
satisfies the Boltzmann equation,
\beq
\deriv{f_p}{t} 
  = \frac{\partial f_p}{\partial t} 
   + {\bm v}_p\cdot \frac{\partial f_p}{\partial{\bm x}}
   + {\bm F}_{\it ext}  \cdot \frac{\partial f_p}{\partial{\bm p}} 
  = C[f_p] ,
\eeq
where ${\bm F}_{\it ext}$ is an external force and $C[f]$ is the 
collision term. In the case of binary scattering the collision integral 
is given by
\beq
C[f_p] = \frac{1}{2 E_p} \int_{k,k',p'} (2\pi)^4 \de(P+K-K'-P') 
  |{\cal M}|^2  D_{2\leftrightarrow 2} ,
\eeq
where 
\beq
\int_q = \int \frac{d^3 q}{(2\pi)^3\,2E_q} 
\eeq
is the one-particle phase space integral, and $D_{2\leftrightarrow2}
=f_pf_k(1+f_{p'})(1+f_{k'}) - (1+f_p)(1+f_k)f_{p'}f_{k'}$ is the 
term involving the distribution 
functions for $2\leftrightarrow2$ scattering. Linearizing in $\de f$ 
we can write $D_{2\leftrightarrow2} = D_0 + \de D$ where
\beq
\de D = -\frac{\f f_k^0 (1+f_{k'}^0) (1+f_{p'}^0)}{T^3} 
  {\bm\De}_g \cdot {\bm\nabla} T,
\eeq
and
\beq 
{\bm\De}_g 
  = g(p) {\bm p} + g(k) {\bm k} - g(k'){\bm k'} - g(p') {\bm p'} \, .
\eeq
Note that the collision integral will vanish for $D=D_0$. This is 
a consequence of detailed balance, which implies that $|{\cal M}|^2$
is invariant under $(p,k)\leftrightarrow(p'\,k')$ and that  $f_p^0f_k^0 
(1+f_{k'}^0) (1+f_{p'}^0) = (1+f_p^0)(1+ f_k^0) f_{k'}^0 f_{p'}^0$.
We can then write the collision integral as
\beq
C[f_p] \equiv -\frac{{\bm F} \cdot {\bm\nabla} T}{T^2}\, , 
\label{beq_rhs}
\eeq
where
\beq
{\bm F} = \frac{1}{2 E_p} \int d\Ga_{k,k',p'} 
  \frac{\f f_k^0 (1+f_{k'}^0) (1+f_{p'}^0)}{T} {\bm \De}_g 
\label{F}
\eeq
and
\beq
\int d\Ga_{k,k',p'} = \int_{k,k',p'} (2 \pi)^4 \de(P+K-P'-K')\, 
|{\cal M}|^2\, .
\eeq
We can simplify the left-hand side of the Boltzmann equation using
our definition of the distribution function in \Eqn{dist}. In general,
we can write
\beq
\deriv f t = -\al_p \frac{\f(1+\f)}{T^2} 
   {\bm p}\cdot{\bm \nabla} T \, , 
\label{alpha_def}
\eeq
where the form of $\al_p$ depends on the properties of the 
quasi-particles we are studying. In the following we will consider 
two cases. The first is phonons with a non-linear dispersion 
relation
\beq 
E_p^{(P)} = v p \left( 1+\gamma p^2\right) 
   \equiv v p \left(1+\eps \frac{v^2\,p^2}{T^2}\right)\, , 
\label{Hdisp}
\eeq
where $v=1/\sqrt{3}$ is the speed of sound, $\gamma$ controls 
the curvature of the dispersion relation, and $\eps\equiv
\gamma T^2/v^2$ is a dimensionless measure of the non-linearity. 
The parameter $\gamma$ was computed by Zarembo \cite{Zarembo:2000pj},
who finds $\gamma = -11/(540\,\De^2)$, where $\Delta$ is the gap in 
the fermion spectrum. The second case is massive kaons with 
\beq
E_p^{(K)} = \sqrt{v_K^2 p^2 + m_K^2} 
  \simeq \frac{v_k^2 p^2}{2 m_K} + m_K \, , 
\eeq
where $v_K=v$. We will specify the kaon mass $m_K$ in Section 
\ref{sec_kaon}. The quasi-particle velocities, $v_p = \pderiv {E_p}{p}$, 
are given by
\beq
v_p^{(P)} = v \left(1+3\eps \frac{v^2 p^2}{T^2}\right)\, , 
\hspace{0.5cm}
v_p^{(K)} =   \frac{v^2 p}{m_K}\, .
\label{vp}
\eeq
The explicit form of $\alpha_p$ is derived in Appendix \ref{appA}.
We find
\beq
\al_p^{(P)} = 4\,\eps\,v^2\, 
    \left(\frac{v^2 p^2}{T^2} -  \frac{20 \pi^2}{7}\right)\, , 
\hspace{0.5cm}
\al_p^{(K)} = \frac{v^4p^2}{2 m_K^2} - \frac{5\,v^2\,T}{2m_K}.
\eeq
We have now expressed the LHS of the Boltzmann equation, \Eqn{alpha_def}, 
in terms of the function $\alpha_p$. The RHS of the Boltzmann equation, 
\Eqn{beq_rhs}, is given in terms of the unknown function $g(p)$. Once 
$g(p)$ is determined the thermal conductivity is given by \Eqn{kappa}.
In practice we will solve the Boltzmann equation using a variational 
procedure. This procedure is based on an expression for the thermal 
conductivity in terms of a suitable integral over the collision term, 
which we will now derive. Consider the following integral over the 
LHS of the Boltzmann equation,
\beq
\frac{c}{T}\int d\Ga\, \deriv f t g(p) {\bm p} 
  = -\frac{c{\bm\nabla} T}{3\,T^3}
      \int d\Ga\, \f (1+\f) p^2 \al_p g(p),
\label{match_c1}
\eeq
where $c$ is a constant. We observe that the RHS of this equation
has the same structure as \Eqn{heat} for the heat flux. For the
two expressions to be equal, we need
\beq
c \int d\Ga\, \f (1+\f) p^2 \al_p g(p) 
          = \int d\Ga\, \f(1+\f) p^2\frac{v_p\,E_p}{p}\, g(p)\, . 
\label{match_c}
\eeq
The two sides of this equation can be matched by using the constraint 
\Eqn{const3}. The constraint implies that constant terms in $v_p E_p/p$ 
and $\al_p$ do not contribute to \Eqn{match_c}. We can then determine
$c$ by matching the $p^2$ terms in $v_p E_p/p$ and $\al_p$. We find
$c = 1$ for both phonons and kaons.  Using this fact 
together with the Boltzmann equation we can express the heat 
current in terms of the collision integral
\beq
{\bm q} = -\frac{{\bm \nabla}T}{3T^3} \int d\Ga\, g(p)\, 
      {\bm p} \cdot {\bm F}\, .
\eeq
The corresponding expression for the thermal conductivity is
\beq
\ka  = \frac{1}{12T^4} \int d\Ga_{p,k,k',p'}\, \De_g^2 \, .
\label{kappa2}
\eeq
where
\beq
\int d\Ga_{p,k,k',p'} = \int_{pkp'k'}
(2\pi)^4 \de(P+K-K'-P') |{\cal M}|^2 \f f_k^0 (1+f_{k'}^0) (1+f_{p'}^0)
\eeq
and we have used \Eqn{F} and the symmetries of the matrix elements to 
derive \Eqn{kappa2}. We now have two expressions for the thermal 
conductivity, \Eqn{kappa} and \Eqn{kappa2}. Both equations depend
on the unknown function $g(p)$, but the equivalence of the two 
equations rests on the fact that $g(p)$ satisfies the Boltzmann 
equation. As we will now show, this fact can be used to determine
$g(p)$. 

\section{Variational Solution to the Boltzmann Equation}
\label{varsol}

 We will solve the Boltzmann equation by expanding $g(p)$ in a 
basis of orthogonal polynomials. The procedure is variational 
in the sense that the result for $\kappa$ obtained in a truncated
basis provides an upper bound on $\kappa$ for the exact solution.  
The expansion has the form 
\beq
g(p) = \sum_s b_s B_s(p^2)\, , 
\label{g_exp}
\eeq
where $B_s(p^2)$ is a polynomial in $p^2$ of order $s$. The coefficient
of the highest power is set to one. This means that $B_0 = 1$, $B_1 = 
p^2 + c_{10}$, etc.  The polynomials $B_s(p^2)$ are orthogonal with
regard to the inner product
\beq
\int d\Ga \f(1+\f) p^2 B_s(p^2) B_t(p^2) \equiv A_s \de_{st} \, .
\label{Bnorm}
\eeq
The functions $B_s(p^2)$ are a generalization of the Laguerre 
polynomials used in solutions of the linearized Boltzmann equation 
in classical physics \cite{Landau:kin}. Starting with $B_0 = 1$, 
we can solve for all higher polynomials and their normalizations, 
$A_s$. This is laid out in more detail in Appendix \ref{app:poly}.  

 Inserting the expansion \Eqn{g_exp} into the constraint \Eqn{const3}
we get 
\beq
0 =
    \int d\Ga \f(1+\f) p^2 B_s(p^2) B_0(p^2)\,,  
\eeq
where we have used $B_0=1$. We conclude that the constraint is 
satisfied if $b_0=0$. Using the polynomial expansion in the first 
expression for $\ka$, \Eqn{kappa}, gives
\beq
\ka 
   = \frac{1}{3T^3}\sum_{s\neq 0} b_s \int d\Ga \f(1+\f)\, p^2\, B_s(p^2)\,
        (a_0 B_0 + a_1 B_1) 
   = \frac{a_1}{3T^3} b_1 A_1 \ , 
\label{kappa_ans}
\eeq
where we have written $v_p E_p/p=a_0B_0+a_1B_1$. The coefficients $a_0$ and 
$a_1$ are determined by \Eqn{vp}. We get $a_1 = 4\eps v^4/T^2$ for 
phonons and $a_1=v^4/2m_K^2$ for kaons. Substituting the polynomial 
expansion into the second expression for the thermal  conductivity, 
\Eqn{kappa2}, gives
\beq
\ka = \frac{1}{12T^4}\sum_{s,t\neq 0} b_s b_t M_{st},
\label{kappa_ans_2}
\eeq
where
\beq
\label{M_st}
M_{st} =  \int d\Ga_{p,k,k',p'} {\bm Q}_s \cdot {\bm Q}_t\, , 
\hspace{0.5cm}
{\bm Q}_s=  B_s(p^2) {\bm p} + B_s(k^2) {\bm k} 
              - B_s(k'^2) {\bm k'} - B_s(p'^2) {\bm p'}\, . 
\eeq
Requiring \Eqn{kappa_ans} and \Eqn{kappa_ans_2} to be 
equal gives an equation for $b_s$
\beq
\frac{a_1}{3T^3} \sum_{s\neq 0}b_s A_s \de_{s1} = \frac{1}{12T^4}
\sum_{s,t\neq 0} b_s b_t M_{st},
\eeq
which is equivalent to the linear equation
\beq
\sum_{t\neq 0} M_{st} b_t = \frac{4a_1T}{3} A_1 \de_{s1}.
\eeq
This equation can be solved by inverting matrix $M$. We get 
\beq
\left(\begin{array}{l} 
    b_1 \\ b_2 \\ \vdots \end{array}\right)
 = \frac{4a_1T}{3} A_1 M^{-1} 
\left(\begin{array}{l} 1 \\ 0 \\ \vdots \end{array}\right)
\eeq 
This equation determines $b_1$ and, using \eqn{kappa_ans},
the thermal conductivity.
We find
\beq
\ka = \left(\frac{4a_1^2}{9 T^2}\right) A_1^2 M^{-1}_{11},
\eeq
where $M_{11}^{-1}$ is the $(1,1)$-element of the matrix $M$, the 
constants $a_1$ are given below \Eqn{kappa_ans} and $A_1$ is given in 
Appendix \ref{appA}. $M$ is an infinite 
matrix, and in practice we solve for $\kappa$ by restricting the 
dimension of the matrix to a finite number $N$. This procedure 
can be shown to be variational in the sense that \cite{Jensen:1969}
\beq
\ka \geq \left(\frac{4a_1^2}{9 T^2}\right) \frac{(b_1 A_1)^2}
{\sum_{s,t\neq 0}^N b_s b_t M_{st}}
\eeq
for any value of $N$.  This right hand side can be written as
\beq
\ka \geq\left(\frac{4a_1^2}{9 c T^2}\right) A_1^2 M^{-1}_{11} \, , 
\label{kappafinal}
\eeq
where $M^{-1}$ is the the inverse of the truncated $N\times N$ 
matrix. The bound is saturated as $N\rightarrow \infty$. 

\section{Collision Terms}
\label{sec_coll}

 The matrix $M$ depends on the $2\leftrightarrow 2$ scattering 
amplitudes. In this Section we will compute the phonon and kaon 
scattering amplitude using a low energy effective lagrangian 
for the CFL phase. 

\subsection{Superfluid Phonon}
\label{sec_phonon}

\begin{figure}
\includegraphics[width=0.8\textwidth]{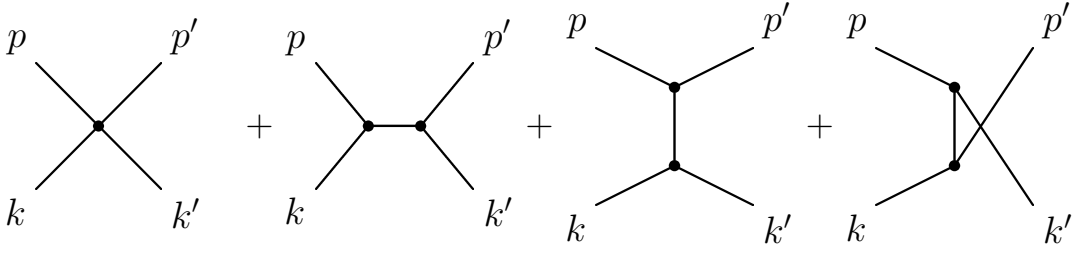}
\caption{Lowest order Feynman diagrams for the 2-phonon 
scattering amplitude. 
\label{feyn}}
\end{figure}

The effective lagrangian for the phonon field $\phi$ is given 
by \cite{Son:2002zn}
\beq
{\cal L} = \half (\p_0 \phi)^2 - \half v^2 (\p_i \phi)^2 - 
  \frac{\pi}{9\mu_q^2} \p_0 \phi(\p_\mu \phi \p^\mu \phi) 
+ \frac{\pi^2}{108\mu_q^4}(\p_\mu \phi \p^\mu \phi)^2 + \ldots \, , 
\eeq
where $v=1/\sqrt{3}$ is the speed of sound and $\mu_q$ is the quark
chemical potential. We have displayed the leading three and four-phonon 
vertices. Higher order terms include higher powers of $\phi$ or 
additional derivatives. These terms are suppressed by powers 
of the typical momentum over the quark chemical potential. The 
speed of sound as well as the coefficients of the three and 
four-phonon vertices are given to leading order in the strong 
coupling constant. The lowest order diagrams that contribute to
phonon-phonon scattering are shown in Figure \ref{feyn}.

The corresponding matrix elements were previously computed in 
connection with the phonon contribution to the shear viscosity of 
the CFL phase \cite{Manuel:2004iv}. The matrix element ${\cal M}
(P,K;P'K')$ where $P,K$ and $P',K'$ are the in and out-going 
four momenta can be written as the sum of the contact term
${\cal M}_c$ and the $s,t,u$-channel phonon exchange diagrams
${\cal M}_{s,t,u}$. The individual terms are 
\bea
i{\cal M}_c &=& \la\left[(P\cdot K)(P'\cdot K) + (P\cdot K')(P'\cdot K) 
     + (P\cdot P')(K\cdot K')\right] \nn
i{\cal M}_s &=& g^2\left[2(p_0+k_0)P\cdot K + p_0 K^2 + k_0 P^2\right]
               \left[2(p'_0+k'_0)P'\cdot K' + p'_0 K'^2 + k'_0 P'^2\right] 
                G(P+K)\nn
i{\cal M}_t &=& i{\cal M}_s (P\leftrightarrow K') \nn
i{\cal M}_u &=& i{\cal M}_t (P\leftrightarrow K), 
\eea
where $G(Q)$ is the phonon propagator and the coupling constants $\la$ 
and $g$ are given by 
\beq
\la = \frac{4\pi^2}{108\mu_q^4}\, ,  \qquad 
g = \frac{2\pi}{9\mu_q^2}.
\eeq
At leading order the phonon propagator is given by $G(Q)=(q_0^2-
E_q^2)^{-1}$ with $E_q=vq$. The thermal conductivity is sensitive 
to non-linearities in the dispersion relation and we will use $E_q$ 
as given in \Eqn{Hdisp}. We note that non-linear effects are 
generated by higher derivative terms in the effective lagrangian. 
Except for the non-linearities in the dispersion relation the 
structure of these terms has not been determined. We have 
verified that the matrix elements $M_{st}$ are not sensitive to 
higher order corrections in the vertices. We also note that 
a non-linear dispersion (with $\eps < 0$) regulates a possible 
on-shell divergence of the matrix element, and as consequence there
is no need to include self-energy corrections in the fermion 
propagator as in \cite{Manuel:2004iv,Rupak:2007vp}.

  We can now insert the scattering amplitude into \Eqn{M_st} and 
compute the matrix elements $M_{st}$. Using the energy and momentum
conserving delta functions as well as spherical symmetry the 12-dimensional 
phase space integral in \Eqn{M_st} can be reduced to a 5-dimensional
integral. First, we use the momentum conserving delta function to 
integrate over $\bm{p'}$. We can choose ${\bm p}$ to lie along the 
$z$-direction and perform the corresponding angular integrals. The 
energy conserving part of the delta function can be used to fix the 
magnitude of ${\bm k'}$.  Finally, we note that the integrand 
only depends on the relative azimuthal angle of ${\bm k}$ and 
${\bm k'}$ and we can trivially integrate over the other azimuthal
angle. The remaining 5-dimensional integral is computed numerically
using standard methods, such as the VEGAS routine \cite{Lepage:1980dq}.  

\subsection{Massive Kaon}
\label{sec_kaon}

 The effective lagrangian for the meson octet is given by 
\cite{Casalbuoni:1999wu,Son:1999cm,Bedaque:2001je}
\beq
 {\cal L} = \frac{f_\pi^2}{4}\Tr\left[\nabla_0\Si\,\nabla_0\Si^\dag 
           - v^2 \nabla \Si\,\nabla \Si^\dag\right] 
   + \frac{af_\pi^2}{2}\Tr\left[M^{-1} |M|(\Si+\Si^\dag)\right]\,
\label{Leff}
\eeq
where $\Si = \exp(i\th^a t^a/f_\pi)$ is the chiral field, 
$t^a$ with $a=1,\ldots, 8$ are the $SU(3)$ Gell-Mann matrices,
$f_\pi$ is the pion decay constant, $M={\rm diag}(m_u,m_d,m_s)$ 
is the quark mass matrix, and $|M|$ is the determinant of the 
mass matrix. The covariant derivative of the chiral field
is given by $\nabla_0\Si = \p_0 \Si - i[A,\Si]$ where $A$ is 
the effective chemical potential $A =-M^2/(2 \mu_q)$. At 
asymptotically high density the constants $f_\pi$, $v$, and 
$a$ can be determined by matching the effective theory to 
perturbative QCD \cite{Son:1999cm,Bedaque:2001je,Schafer:2001za}
\beq
 f_\pi^2= \frac{21-8\ln2}{18}\frac{\mu^2}{2\pi^2}\, , \qquad \qquad 
 v =\frac{1}{\sqrt{3}}\, , \qquad \qquad
 a = \frac{3 \De^2}{\pi^2 f_\pi^2}\, , 
\eeq
where $\De$ is the fermionic energy gap at zero temperature.
For $m_s\gg m_u,m_d$ the lightest excitations in the theory 
are kaons. The leading terms in the effective theory involving
kaons are ${\cal L}_0+{\cal L}_{\it int}$ with 
\bea
{\cal L}_{0} 
     &=& \frac{1}{2}(\partial_\mu K^0)(\partial^\mu \bar{K}^0) 
        +\frac{1}{2}(\partial_\mu K^+)(\partial^\mu K^-)
        -\frac{1}{2} m_{K^0}^2 K^0 \bar{K}^0 
        -\frac{1}{2} m_{K^+}^2 K^+ K^- \, , \nn
{\cal L}_{\it int} 
     &=& -\frac{1}{24}\left[\la_0 \left(K^0 \bar{K^0}\right)^2 
         + \la_+ \left(K^+ K^-\right)^2
         + \left(\la_0 + \la_+\right) 
            \left(K^0\bar{K^0}\right)\left(K^+ K^-\right)\right],
\label{l_kaon}
\eea
where $\la_{0,+} = m_{K^{0,+}}^2/f_\pi^2$ and
\beq
m_{K^0}^2 = am_u(m_d+m_s)\, , \qquad m_{K^+}^2 = am_d(m_u+m_s)\, .
\eeq
We have not included kaon interactions that involve derivatives
of the kaon field. These terms arise from expanding the first 
term in \eqn{Leff}, see \cite{Alford:2007qa}. For energies and
momenta that are small compared to the kaon mass these terms 
are suppressed as compared to the terms in \Eqn{l_kaon}.

\begin{figure}
\includegraphics[scale = 1.25]{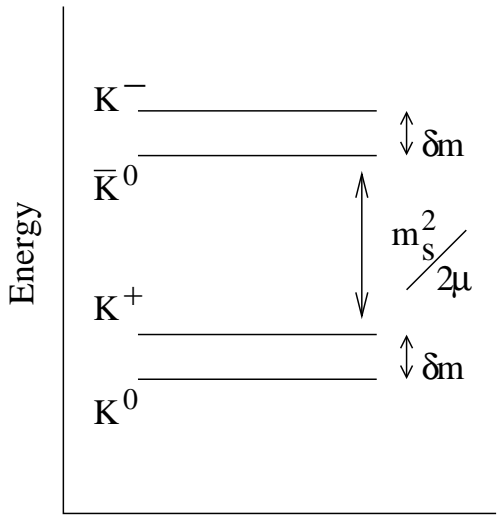}
\caption{Spectrum of energies for the charged and neutral kaons.  The
mass splitting between $K^0$ and $K^+$ is given by $\de m$ and
the larger splitting between the kaons containing ${\bar s}$ and those
containing $s$ is given by $m_s^2/2\mu_q$.  \label{kaon_energy}}
\end{figure}

  The spectrum of kaons is shown schematically in Figure \ref{kaon_energy}.
The mass difference between the isospin doublets $(K^0,K^+)$ and
$(\bar{K}^0,K^-)$ is proportional to $m_s^2/(2\mu_q)$, while the 
mass splitting with the doublets is  $\de m^2 = am_s(m_d-m_u)$.
In the following we will only include the lightest kaon, the $K^0$. 
The relevant interaction term is 
\beq
{\cal L}_{int} = -\frac{m_{K^0}^2}{24 f_\pi^2}(K^0 \bar{K^0})^2\, ,
\eeq
and the corresponding scattering amplitude is 
\beq
|{\cal M}|^2 = \frac{m_{K^0}^4}{f_\pi^4}\, .
\eeq
This expression can be inserted into \Eqn{M_st} in order to compute 
the matrix $M_{st}$. As in the case of phonon scattering, the 
phase space integral can be reduced to a 5-dimensional integral 
over the magnitudes of ${\bm p}$ and ${\bm k}$, the polar angles 
$\cos(\th_k)$ and $\cos(\th_{k'})$ describing the angles between 
${\bm p}$ and the respective momenta, and one azimuthal angle.
  
\section{Results}
\label{sec_res}

  Before we show numerical results for the thermal conductivity, we 
would like to discuss the dependence of $\kappa$ on the dimensionful
parameters that appear in the problem, the temperature $T$, the 
quark chemical potential $\mu_q$, the fermion gap $\Delta$, and
the mass of the kaon. We begin by studying the scaling of the 
matrix elements $M_{st}$. In the phonon case, we can rescale all 
momenta by the temperature and write $M_{st}$ as a product of 
the dimensionful parameters and a dimensionless phase space 
integral. A subtlety arise due to the possible role of on-shell
divergences. There is no on-shell sensitivity in the phonon 
contribution to the shear viscosity and the corresponding matrix 
elements exhibit the naive scaling behavior, but the phonon mean 
free path is sensitive to the phonon self energy near the on-shell 
point \cite{Rupak:2007vp,Manuel:2004iv}. We find that the thermal 
conductivity also depends on the behavior of the phonon propagator 
near the on-shell point. In the present calculation this behavior 
is governed by non-linearities in the dispersion relation. 
Consider the contribution to $M_{st}$ from nearly on-shell 
phonons in the $s,t$ or $u$-channel. We find
\beq
M_{st}^{(P)} \,\sim\, \frac{T^{2s+2t+14}}{\mu^8}\int  
       \frac{dq^2d\Gamma'}{(q^2 + \eps f(q,\Ga'))^2} 
\,\sim\, \frac{T^{2s+2t+14}}{\mu^8}\frac{1}{|\eps|}
\,\sim\, \frac{T^{2s+2t+12}\De^2}{\mu^8}
\label{M_st_ph_scal}
\eeq
where $q$ represent the momentum transfer in the relevant channel,
$d\Gamma'$ is a phase space integral over the remaining momenta, 
and $f(q,\Ga')$ is a function of these variables. 

\begin{figure}
\includegraphics[width=0.6\textwidth]{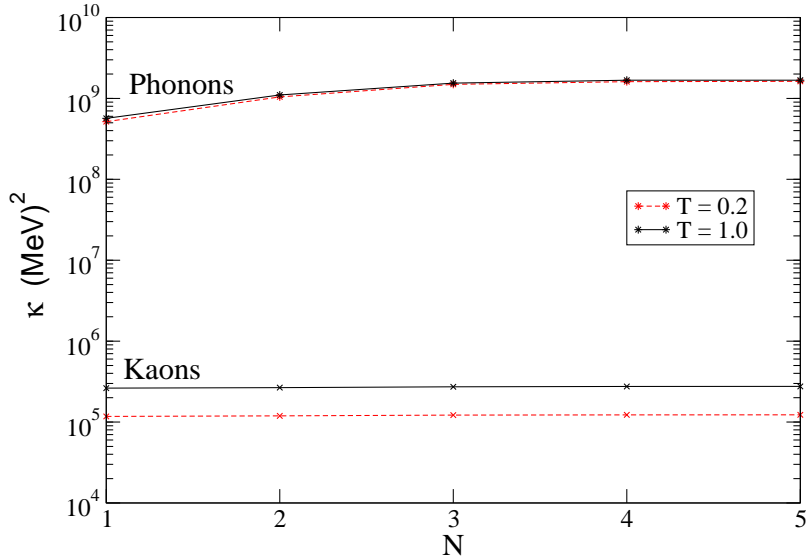}
\caption{Numerical results for the thermal conductivity of the 
CFL phase due to phonons and kaons as a function of the basis 
size $N$ used in the variational procedure. Results are shown 
for two different temperatures $T=0.2$ MeV and $T=1$ MeV. We have 
used $\mu = 400~\MeV$, $\De = 100~\MeV$ for the phonons and $m_K = 
10~\MeV$, $\mu_K = 5~\MeV$ and $f_\pi = 100~\MeV$ for the kaons.
Numerical results are shown as stars and crosses, the lines are 
only serve to guide the eye.  
\label{thermN}}
\end{figure}

 In the kaon case we can assume that the temperature is smaller
than all the other parameters and write the dispersion as $E_q 
= m_K + q^2/(2m_K)$.  We rescale the momenta by $\sqrt{m_K T}$ and
extract all dimensionful parameters from the phase space integral.  
We get
\beq
 M_{st}^{(K)} \,\sim\,  \frac{\la_0^2}{m_K^3} (m_K T)^{s+t+9/2} e^{-2\de m/T} 
  \,\sim\, \frac{m_K}{f_\pi^4} (m_K T)^{s+t+9/2} e^{-2\de m/T} 
\label{M_st_K_scal}
\eeq
Using \Eqn{M_st_ph_scal} and \Eqn{M_st_K_scal}, as well as the 
scaling of the normalization factor $A_1$ derived in Appendix 
\ref{app:poly}, we find
\beq
\ka \sim \left\{
 \begin{array}{ll}
   \frac{\mu_q^8}{\De^6}                      & \; {\rm phonons}, \\ 
   \frac{f_\pi^4}{m_K^2}\sqrt{\frac{T}{m_K}}  & \; {\rm kaons}.    
\end{array}\right.
\label{kpar}
\eeq
We note that the contribution to the thermal conductivity from
phonons is independent of temperature.  The naive power counting 
gives $\ka \sim \mu^8/T^6$, but the contribution from nearly 
on-shell phonons modifies the naive scaling behavior by a factor 
$\eps^3$.  Two powers of $\eps$ come from the $a_1^2$ term and 
one power of $\eps$ comes from the nonlinear dispersion cutting 
off the collinear regimes in the propagators. Using $\epsilon\sim
T^2$ then leads to a thermal conductivity that is independent 
of temperature. 

 The thermal conductivity due to kaons is not exponentially 
suppressed by $\exp(-m_K/T)$ as one might naively expect, because
the exponential factors in $A_1^2$ and $M_{11}$ cancel at leading 
order. This result is analogous to the well known fact that the 
transport properties of a dilute gas are independent of the density
of the gas. We should note, however, that in practice the range 
of validity of this result is limited to temperatures that are 
not too small. At very small temperature the kaon mean free path
is bigger than the system size, and heat transport is no longer 
a diffusive process. 

\begin{figure}
\includegraphics[width=0.6\textwidth]{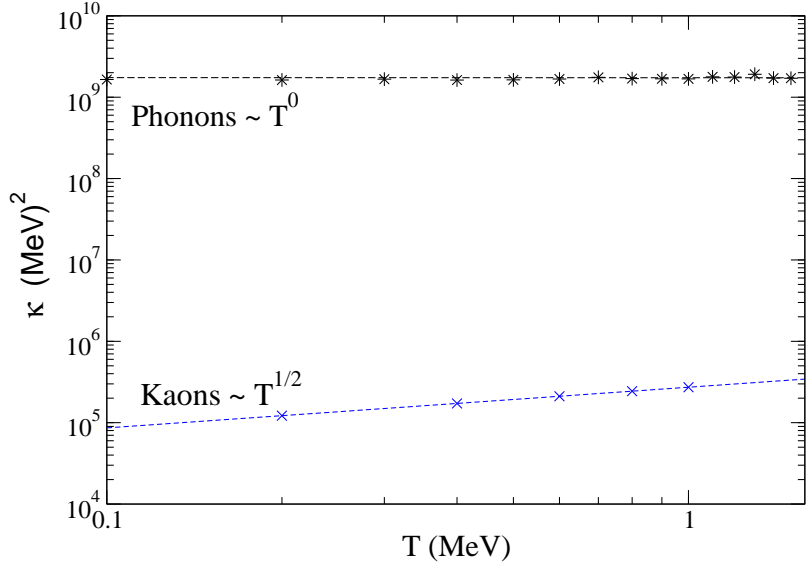}
\caption{Numerical results for the contribution of phonons and 
kaons to the thermal conductivity of the CFL as a function of 
temperature. The results shown in this figure were obtained using 
the variational method with $N=5$. We have used $\mu = 400~\MeV$, 
$\De = 100~\MeV$  for the phonons and $m_K = 10~\MeV$, $\mu_K = 5~\MeV$ 
and $f_\pi = 100~\MeV$ for the kaons.
\label{thermT}}
\end{figure}
 
 Numerical results for the thermal conductivity are shown in 
Figs.~\ref{thermN} and \ref{thermT}. In Fig.~\ref{thermN} we 
show the dependence of $\kappa$ on the number of basis functions. 
We observe that the convergence in the case of the kaon contribution 
is excellent -- using $N=1,2$ gives results within a few percent of 
the final result.  Also, the rate of convergence is independent of 
the temperature. In the case of the phonon contribution the rate 
of convergence is slower, but going up to $N=5$ appears to give 
converged results. 

 Fig.~\ref{thermT} shows the thermal conductivity as a function of 
temperature. We have used the results for $N=5$. We observe that 
thermal conductivity due to kaons follows the square root dependence 
predicted by \Eqn{kpar}. Also, the thermal conductivity due to 
phonons is approximately temperature independent, in agreement 
with \Eqn{kpar}. The numerical results are well represented by 
\beq
\ka^{(P)} \gtrsim 4.01 \times 10^{-2}\, \frac{\mu^8}{\De^6}\,\MeV^2\, ,  
\qquad \qquad
\ka^{(K)} \simeq 0.86\, \frac{f_\pi^4}{m_K^2}\sqrt{\frac{T}{m_K}}~\MeV^2\, . 
\eeq
We observe that for $m_K$ in the few MeV range the phonon contribution
is significantly larger than the kaon contribution.

\section{Conclusions}
\label{sec_concl}

 We can convert these result into CGS units. We get 
\beq
\ka^{(P)} \gtrsim 1.04 \times 10^{26}\, \frac{\mu_{500}^8}{\De_{50}^6}\, 
\frac{{\rm erg}}{\rm cm\, s\, K} \, , 
 \qquad 
\ka^{(K)} \simeq 2.81 \times 10^{21}\, f_{\pi,100}^4\, T_{\MeV}^{1/2}\, 
m_{10}^{-5/2}\, \frac{{\rm erg}}{\rm cm\, s\, K}\, , 
\eeq
where $\mu_{500}$ is the quark chemical potential in units of $500~\MeV$,  
$\De_{50}$ is the gap in units of $50~\MeV$, $f_{\pi,100}$ is the pion
decay constant in units of $100~\MeV$, $T_{\MeV}$ is the temperatures
in units of $\MeV$ and $m_{10}$ is the kaon mass in units of $10~\MeV$.

 These numbers can be compared to a number of results in the literature.
The thermal conductivity of the CFL phase was previously studied by 
Shovkovy and Ellis \cite{Shovkovy:2002kv}. Based on a mean free path 
estimate they find that the contribution due to phonons is 
$\ka^{(P)} \simeq 0.7 \times 10^{32}\, T_{\MeV}^3 R_{0,\rm{km}}\,
\frac{{\rm erg}}{\rm cm\, s\, K}$, where $R_{0,\rm{km}}$ is the 
phonon mean free path in units of 1 km. This particular number 
was chosen because the authors argued that the mean free path 
is so long that it is effectively cut off by the size of the quark 
matter core. In the present work we have demonstrated that in the 
framework of kinetic theory there is no simple connection between 
the phonon mean free path and the thermal conductivity. We note, 
however, that if one replaces the mean free path in the Shovkovy-Ellis 
estimate with a more accurate result from kinetic theory, which is 
on the order of millimeters rather than kilometers \cite{Manuel:2004iv}, 
one obtains a thermal conductivity which is close to our result.
Shovkovy and Ellis also argued that there is a large contribution 
to $\kappa$ from photons. The photon mean free path is indeed on 
the order of the size of the star \cite{Jaikumar:2002vg}. This 
implies, however, that photons are not thermally coupled to the 
a CFL quark matter core. 

 The thermal conductivity of unpaired quark matter at low temperature 
is $\kappa=0.5 m_D^2/\alpha_s^2$ \cite{Heiselberg:1993cr}, where 
$m_D$ is the gluon screening mass and $\alpha_s$ is the strong 
coupling constant. Using $m_D\simeq 500$ MeV and $\alpha_s=0.3$ 
this gives $\kappa\simeq 10^4\, {\rm MeV}^2$, about 5 orders
of magnitude larger than the thermal conductivity of the CFL 
phase. The thermal conductivity of nuclear matter was studied in 
\cite{Flowers:1976,Flowers:1979,Itoh:1983,Wambach:1992ik,Baiko:2001cj}.  
A typical value at nuclear matter saturation density and $T\simeq 1$ 
MeV is $\kappa\simeq (10^{20}-10^{21})\frac{{\rm erg}}{\rm cm\, s\, K}$, 
which is about 5 orders of magnitude smaller than the thermal 
conductivity of the CFL phase.

 We can convert the thermal conductivity to a timescale for thermal
diffusion. The thermal diffusion constant is $\chi=\kappa/c_V$, 
where $c_V$ is the specific heat. The time scale for thermal 
diffusion over a distance $R$ is 
\beq
\tau \simeq \frac{c_V R^2}{\ka} \, .
\eeq
Using $c_V=2\pi^2 T^3/(15v^3)$ for phonons, as well as $T=1$ MeV, 
$\mu =500$ MeV, $\Delta = 50$ MeV, and $R=1$ km we get $\tau\simeq 
10\,{\rm s}$, indicating that a CFL quark core will become isothermal rapidly. 

 In this paper we have studied the thermal conductivity of the CFL 
phase using a linearized Boltzmann equation including a collision 
term that involves $2\leftrightarrow 2$ scattering between phonons 
or kaons only. We have not studied a coupled transport equation for 
phonon-kaon scattering. Given that the relaxation times of phonons 
and kaons are quite different, it should be possible to find approximate 
solutions to the coupled problem. We have only provided a simple
estimate for the relaxation time of temperature gradients. More 
accurate estimates will require a detailed model of the initial 
temperature profile, and a full solution of the dissipative 
two-fluid hydrodynamic equations for a CFL superfluid.

Acknowledgments:  We would like to thank C. Manuel and
M. Manarelli for very helpful communications.  This work 
was supported in parts by the US Department of Energy grant 
DE-FG02-03ER41260. 

\appendix
\renewcommand{\theequation}{\Alph{section}.\arabic{equation}}
\section{Streaming terms in the Boltzmann equation}
\label{appA}
\subsection{Superfluid Phonon}

 In this appendix we compute the quantity $\alpha_p$, which is 
related to the streaming terms (the left-hand side) of the Boltzmann 
equation. We will use the dispersion relation for the phonon given by
\beq
E_p = v p\left(1+\gamma p^2\right) 
  \equiv v p\left(1+\eps \frac{v^2\,p^2}{T^2}\right)\, . 
\eeq
The left-hand side of the Boltzmann equation can be written as
\beq
\deriv {f_p} t = \pderiv {f_p} t + {\bm v}_p\cdot {\bm \nabla} f_p \, , 
\eeq
where ${\bf v}_p$ is the particle velocity. In local thermal 
equilibrium the time dependence of $f_p$ arises from the time 
dependence of the local temperature and fluid velocity. We have 
\beq
\pderiv {f_p} t 
= -\frac{ \f (1+\f)}{T}\left(\frac{E_p}{T} \pderiv T t 
  + {\bm p}\cdot \pderiv {\bm u} t \right)\ , 
\eeq
where we have used the fact that by going to the local rest frame  
we can set the local fluid velocity (but not its derivatives to zero). 
We have also used that the number of phonons is not conserved, and 
the phonon chemical potential is zero. The spatial derivatives can 
be written as 
\beq
{\bm v}_p\cdot {\bm \nabla} f_p =
   -\frac{ \f (1+\f)}{T}\left(\frac{E_p}{T} {\bm v}_p\cdot {\bm \nabla} T 
  + {\bm v}_p\cdot {\bm \nabla} ({\bm p}\cdot {\bm u}) \right) \, . 
\eeq
We can simplify these expressions using the Euler equation 
\beq
\pderiv {\bm u} t = -\frac{1}{\rho} {\bm \nabla} P 
= -\frac{1}{\rho}\frac{dP}{dT} {\bm \nabla} T
\eeq
where $P$ is the pressure and $\rho$ is the mass density. The mass
density is defined by
\beq 
{\bm \pi} = \rho {\bm u}\, ,
\eeq
where ${\bm \pi}$ is the momentum density of the fluid. We can 
now collect all terms that contain a gradients of $T$.  We find
\beq
\deriv {f_p} t = -\frac{\f (1+\f)}{T}
    \left(\frac{E_p}{T} {\bm v}_p\cdot {\bm \nabla} T  
    - \frac{1}{\rho}\deriv P T {\bm p} \cdot {\bm \nabla} T\right)\, . 
\eeq
Noting that ${\bm v}_p = (v_p/p) {\bm p}$ and using our definition of
$\al_p$ in \Eqn{alpha_def}, we can write
\beq
\al_p = \frac{T}{\rho}\deriv P T - E_p\frac{v_p}{p}\, .
\label{alpha_ph_thermo}
\eeq
The pressure of the phonon gas is given by
\beq
P = -T\int d\Ga\, {\rm ln}\left( 1-e^{-E_p/T} \right), 
\eeq
leading to
\beq
\deriv P T = \frac{1}{3 T} \int d\Ga (3E_p + v_p\,p)\f
  \simeq  \frac{2\pi^2 T^3}{45 v^3}
      \left(1 + \frac{60\eps\pi^2}{7}\right) \, , 
\eeq
where we have kept terms up to linear order in $\eps$. The mass 
density is given by
\beq
\rho  = \frac{1}{3 T} \int d\Ga p^2 \f(1+\f) 
 \simeq \frac{2\pi^2 T^4}{45 v^5}\left(1 + 20\eps\pi^2\right)\, . 
\eeq
Inserting these results into \Eqn{alpha_ph_thermo} we get
\beq
\al_p^{(P)} = 4\eps v^2\,\left[\frac{v^2\,p^2}{T^2} 
    - \frac{20\pi^2}{7}\right]\, . 
\label{app_alpha}
\eeq
We note that $\alpha_p=0$ for $\eps=0$. This result is consistent
with \Eqn{const3}. In the case of phonons with an exactly linear 
dispersion relation, the LHS side (the streaming term) of the 
Boltzmann equation is zero. The RHS (the collision term) is 
zero if $\delta f_p$ is a zero mode of the linearized collision
operator. The zero mode associated with momentum conservation 
satisfies the constraints in \Eqn{del_f_constr} and, because
of \Eqn{const3}, gives a vanishing thermal conductivity. For 
$\eps\neq 0$ both the LHS and the RHS are not zero, and the 
corresponding solution of the linearized Boltzmann equation 
leads to a finite thermal conductivity.

\subsection{Massive Kaon}

 In this section we compute the streaming term for kaons. There 
are two important differences as compared to the phonon case: 
kaons are massive, and they carry conserved charges (hypercharge
and isospin) which couple to chemical potentials. For simplicity
we consider only the neutral kaon, which couples to a single 
chemical potential (a linear combination of isopsin and hypercharge). 
We will also consider the limit of low temperature, such that $T\ll m,
\mu,\de m \equiv m-\mu$, where $m\equiv m_K$ and $\mu\equiv \mu_K$.  
The left-hand side of the Boltzmann equation can be written as
\beq
\frac{df_p}{dt} = \pderiv{f_p}{\bm u}\cdot\deriv{\bm u}{t} 
      + \pderiv{f_p}{ T} \deriv T t
      + \pderiv{f_p}{\mu} \deriv \mu t 
 + {\bm v}_p\cdot {\rm gradient\ terms}
\eeq
The time derivatives of the temperature and the chemical potential can
be converted to spatial derivatives using the the continuity equation. 
These terms do not contribute to the thermal conductivity. The derivative 
of the velocity field can be rewritten using the Euler equation 
\beq
\deriv {\bm u} t = -\frac{1}{\rho} {\bm\nabla} P.
\eeq
Spatial derivatives of the velocity field only contribute to the shear 
viscosity. The spatial derivative of the chemical potential can be 
rewritten using the Gibbs-Duhem relation $dP = n d\mu + s dT$,  where 
$n$ is the density and $s$ is the entropy density. We get
\beq
{\bm \nabla} \mu = \frac{1}{n}{\bm \nabla} P 
      - \frac{s}{n} {\bm \nabla} T\, .
\eeq
Collecting gradients of the temperature and pressure we find
\beq
\deriv{f_p}{t} = \left(\pderiv{f_p}{T} - \frac{s}{n} \pderiv{f_p}{\mu}\right)
  {\bm v}_p \cdot {\bm\nabla} T 
 + \left(\frac{{\bm v}_p}{n} \pderiv{f_p}{\mu} - \frac{1}{\rho}
\pderiv{f_p}{\bm u} \right)\cdot {\bm \nabla} P\, . 
\label{dfdt_mu}
\eeq
At this point we need to specify explicit expressions for the 
thermodynamic quantities. The pressure of an ideal gas of massive 
bosons is given by
\beq
P =  -T\int d\Ga\, {\rm ln}\left(1-e^{-(E_p-\mu)/T}\right)\, . 
\eeq
The entropy density, particle density, and mass density are
\bea
s &=&    
   \frac{1}{3T} \int d\Ga\,  
      \f \left(3\left(E_p - \mu\right) + v_p p\right)
  \simeq \left(\frac{\de m}{T} + \frac{5}{2}\right) n\, , \nn
n &=& 
  \int d\Ga\, \f 
 \simeq \left(\frac{m\,T}{2\pi v^2}\right)^{3/2} e^{-\de m/T}\, ,  \nn
\rho &=& \frac{1}{3T} \int d\Ga\, p^2\, \f(1+\f)
   \simeq \frac{m}{v^2} n\, ,
\eea
where we have given analytical results in the low temperature limit. 
We also need the derivatives of the distribution function with 
respect to the thermodynamic quantities, 
\bea
\pderiv f T       &=& \frac{\f(1+\f)}{T} \frac{E_p-\mu}{T} \nn
\pderiv f \mu     &=& \frac{\f(1+\f)}{T} \nn
\pderiv f {\bm u} &=& \frac{\f(1+\f)}{T} {\bm p}.
\eea
The coefficient of the ${\bm \nabla}P$ term in \Eqn{dfdt_mu} is given
by
\beq
\frac{{\bm v}_p}{n} \pderiv f \mu - \frac{1}{\rho} \pderiv f {\bm u} = 
\frac{\f(1+\f)}{T}\left(\frac{{\bm v}_p}{n} - \frac{\bf m}{\rho}\right).
\eeq
Using $v_p = v^2 p/E_p \sim v^2 p/m$ we can see that $v_p/p \sim v^2/m$. 
We can also show that $n/\rho = v^2/m$. This implies that the coefficient 
of ${\bm \nabla}P$ is zero. The ${\bm \nabla}T$ term is 
\beq
\deriv f t = \left(\pderiv f T - \frac{s}{n} \pderiv f \mu\right)
  {\bm v}_p\cdot {\bm \nabla} T  
 = \frac{\f(1+\f)}{T}\left(\frac{E_p-\mu}{T} - 
\frac{s}{n}\right){\bm v}_p\cdot {\bm \nabla} T\, . 
\eeq
We can then read off the coefficient $\al_p$. We find 
\beq
\al_p^{(K)} = \frac{v^2}{m}\left[E_p - \mu - \frac{Ts}{n}\right] 
\simeq \frac{v^4 p^2}{2m^2} - \frac{5 v^2 T}{2m} \, . 
\eeq

\section{Orthogonal polynomials}
\label{app:poly}

 In this appendix we collect some explicit expressions for the
orthogonal polynomials $B_s(p^2)$ introduced in section \ref{varsol}.  
The starting point is \Eqn{Bnorm}
\beq
\int d\Ga \f(1+\f) p^2 B_s(p^2) B_t(p^2) \equiv A_s \de_{st}\, ,
\eeq
together with $B_0 = 1$.  We write 
\beq
\begin{array}{lll}
B_0 &=& 1  \\
B_1 &=& p^2 + c_{10} \\
B_2 &=& p^4 + c_{22} p^2 + c_{20} \\
B_3 &=& p^6 + c_{34} p^4 + c_{32}p^2 + c_{30} \\
\vdots &&
\end{array}\, , 
\eeq
and iteratively determine the coefficients $c_{st}$. For example, we 
find 
\bea
c_{10} &=& \left\{\begin{array}{ll}
  -\frac{20\pi^2 T^2}{7 v^2}& \;\; {\rm phonons}, \\ 
  -\frac{5 m T}{v^2}        & \;\; {\rm kaons}.
\end{array}\right. 
\eea
which implies that $B_1 \propto \al_p$.  This relation plays a role
in ensuring the consistency of the linearized Boltzmann equation. The
variational function $g(p)=B_0$ is a zero mode of the linearized 
collision operator. We already showed that a term $g(p)\propto B_0$
is eliminated by the constraints, \Eqn{const3}, and that it does not 
contribute to the thermal conductivity. The collision operator acting 
on any function $g(p)$ is orthogonal to $B_0$, because $B_0$ is a zero 
mode, and the collision operator is hermitean. This implies that the 
streaming term must be orthogonal to $B_0$ also. The streaming 
term is proportional to $\alpha_p$, and orthogonality to $B_0$ 
follows from the relation $\alpha_p\propto B_1$. 

 The final result for the thermal conductivity requires the 
normalization constant $A_1$. Using the result for $B_1$ we get 
\beq
A_1 = \int d\Ga\, \f(1+\f) p^2 (p^2 + c_{10})^2 
    = \left\{\begin{array}{ll}
  \frac{256\pi^6}{245 v^9} T^9  & \;\; {\rm phonons}, \\ 
  \frac{15}{32\pi^{3/2}} \left(\frac{2 m T}{v^2}\right)^{9/2} e^{-\de m/T} 
                                & \;\; {\rm kaons}. 
          \end{array}\right. 
\label{A1}
\eeq

\end{document}